\documentclass[%
 reprint,
 amsmath,amssymb,
 aps,
]{revtex4-2}

\usepackage{graphicx}
\usepackage{dcolumn}
\usepackage{bm}

\begin{document}

\preprint{APS/123-QED}

\title{Wide-Angle Invisible Dielectric Metasurface Driven by Transverse Kerker Scattering}

\author{Xia Zhang}%
 \email{xzhang@tcd.ie}
\author{A Louise Bradley}%
 \email{bradlel@tcd.ie}
\affiliation{%
School of Physics, CRANN and AMBER, Trinity College Dublin, Ireland
}%

\date{\today}

\begin{abstract}

Interference is the cornerstone of Huygens source design for reshaping and controlling scattering patterns. The conventional underpinning principle, such as for the Kerker effect, is the interference of electric and magnetic dipole and quadrupole modes.  Here a route to  realize transverse Kerker scattering through employing only the interference between the electric dipole and magnetic quadrupole is demonstrated. The proposed approach is numerically validated in an ultra-thin Silicon square nanoplate metasurface, and is further verified by multipole decomposition. The metasurface is shown to be invisible for near-infrared wavelengths and with an enhanced electric field in the region of the nanoparticle. Additionally, we develop further the proposed approach with practical implementation for invisibility applications by exploring the effects of the aspect ratio of the square plate nanoresonator, the inter-particle separation, and the presence of a substrate. Further it is demonstrated that invisibility can be observed at oblique incidence up to 60$^{\circ}$ for a transverse magnetic plane wave. The results are relevant for Huygens metasurface design for perfect reflectors, invisibility and devices for harmonic generation manipulation. 

\end{abstract}

\maketitle


\section{Introduction}

In 1983, Kerker \emph{et al}. theoretically revealed the possibility of asymmetric scattering by a magneto-dielectric particle \cite{kerker1983electromagnetic}. Suppressed backward scattering and near-zero forward scattering are known as the first Kerker condition and the second Kerker condition, respectively \cite{alu2010does, garcia2010nanoparticles}. The principle underpinning such asymmetric scattering patterns is the interference of electric dipole (ED) and magnetic dipole (MD) resonances, which are provided by the subwavelength particle, called a Mie resonator. High-index dielectric resonators outperform plasmonic counterparts due to the higher intrinsic ohmic losses associated with the latter at optical frequencies \cite{yu2015high,staude2017metamaterial}. Additionally, tailored by geometry, the dielectric resonator can have overlapping ED and MD resonances and provide a platform for light manipulation \cite{geffrin2012magnetic, evlyukhin2012demonstration,fu2013directional,kuznetsov2016optically,kwon2018transmission}.

Arranging the resonators periodically in a planar geometry constitutes a Huygens metasurface under the illumination of an incident electromagnetic (EM) wave \cite{decker2015high,wong2018perfect}. Each resonator behaves as the secondary source, collectively defining the outgoing beam direction \cite{holloway2012overview,lin2014dielectric,chen2016review}. Conceptually similar to Kerker-type scattering of a single nanoresonator, a Mie-type resonant Huygens metasurface exhibits near-unity efficiency functionalities such as perfect reflection and/or perfect transmission under normal \cite{yu2015high, pfeiffer2013metamaterial,kim2014optical,staude2013tailoring,babicheva2017resonant} and oblique light wave incidence  \cite{abujetas2018generalized, nieto2011angle}. Since first reported three decades ago, the Kerker effect has been the cornerstone for exotic scattering pattern reshaping employing the coherent interplay of multipolar modes, including the only recent proposed transverse scattering in dielectric resonators  \cite{lee2018simultaneously,bag2018transverse,shamkhi2019transverse,liu2019lattice,shamkhi2019transparency}. It is a fascinating phenomenon, arising due to the simultaneous fulfillment of the first and second Kerker conditions. The travelling EM wave remains unperturbed for the lossless case, including amplitude and phase. However, remarkably different than a piece of transparent glass, the resonator has concentrated EM field in the transverse plane.  Rather than scattering forward or backward relative to the incident light propagation direction, the scatter, a single resonator or a metasurface, scatters the light only in the transverse plane. In contrast to most Huygens source employing the interplay of only the ED and MD, the coherent contributions from available higher EM modes, the electric quadrupole (EQ) and magnetic quadrupole (MQ), play a pivotal role in transverse scattering. Upon inspection of the Green’s tensor components of all multipole moments contributing to far field scattering  \cite{evlyukhin2010optical,evlyukhin2016optical,terekhov2019multipole}, it is seen that it is fundamentally impossible to realize symmetric scattering through the only interplay of ED and MD modes. The scattered electric field generated by ED or MQ displays an even parity in the plane of incidence while that of the EQ or MD display an odd parity. However, it is theoretically feasible that transverse scattering pattern can be formed via interplay  of only the ED with the MQ or the MD with the EQ. 

In sharp contrast to the work presented in Ref. \cite{shamkhi2019transverse}, which fulfills the generalized amplitude and phase conditions of ED, MD, EQ and MQ, here in this work we propose a different route to realize transverse scattering using the ED and MQ modes only. It could also be theoretically realized by the  interplay of MD and EQ, which has been proposed in Ref. \cite{liu2019lattice} and shown in a proof-of-concept demonstration using core-shell SiO2@InSb and Si@InSb in a 1D metalattice geometry in the Terahertz range. Our proposed method enables the realization of metasurface invisibility in a facile extremely-thin Silicon (Si) square nanoplate metasurface, which is more practically feasible and offers technically easier integration into devices than core shell-type structures. Additionally, the invisibility can be tuned to a particular angle for perfect transparency. Furthermore, the proposed structure can be realized for operation in the visible and near-infrared ranges and paves the way for applications for high-efficiency nonlinear nanophotonics for second harmonic generation \cite{liu2018enhanced,koshelev2020subwavelength}.

Therefore, in this work, we will first explore the principle underpinning the scattered electric field parity within a single nanoparticle case in spherical coordinates and a metasurface case in Cartesian coordinates. The conditions for perfect transverse scattering, which has an even parity with strictly no scattering in the forward and backward directions, are generalized. Additionally, we realize the transverse Kerker scattering in the extremely-thin dielectric metasurface, which shows  dominant ED and MQ modes with negligible MD and EQ modes. The transverse Kerker scattering is validated by numerical simulations including the scattering pattern and electric field map. For practical applications, the dependence of the interplay between the ED and MQ on the aspect ratio of the resonator, the inter-particle separation, the substrate and the angle of incidence   are thoroughly studied. Our study will inspire new metasurface design and practical applications of invisible or transparent metasurface.

\section{Theory}
\begin{figure}
\includegraphics[width=1.0\linewidth]{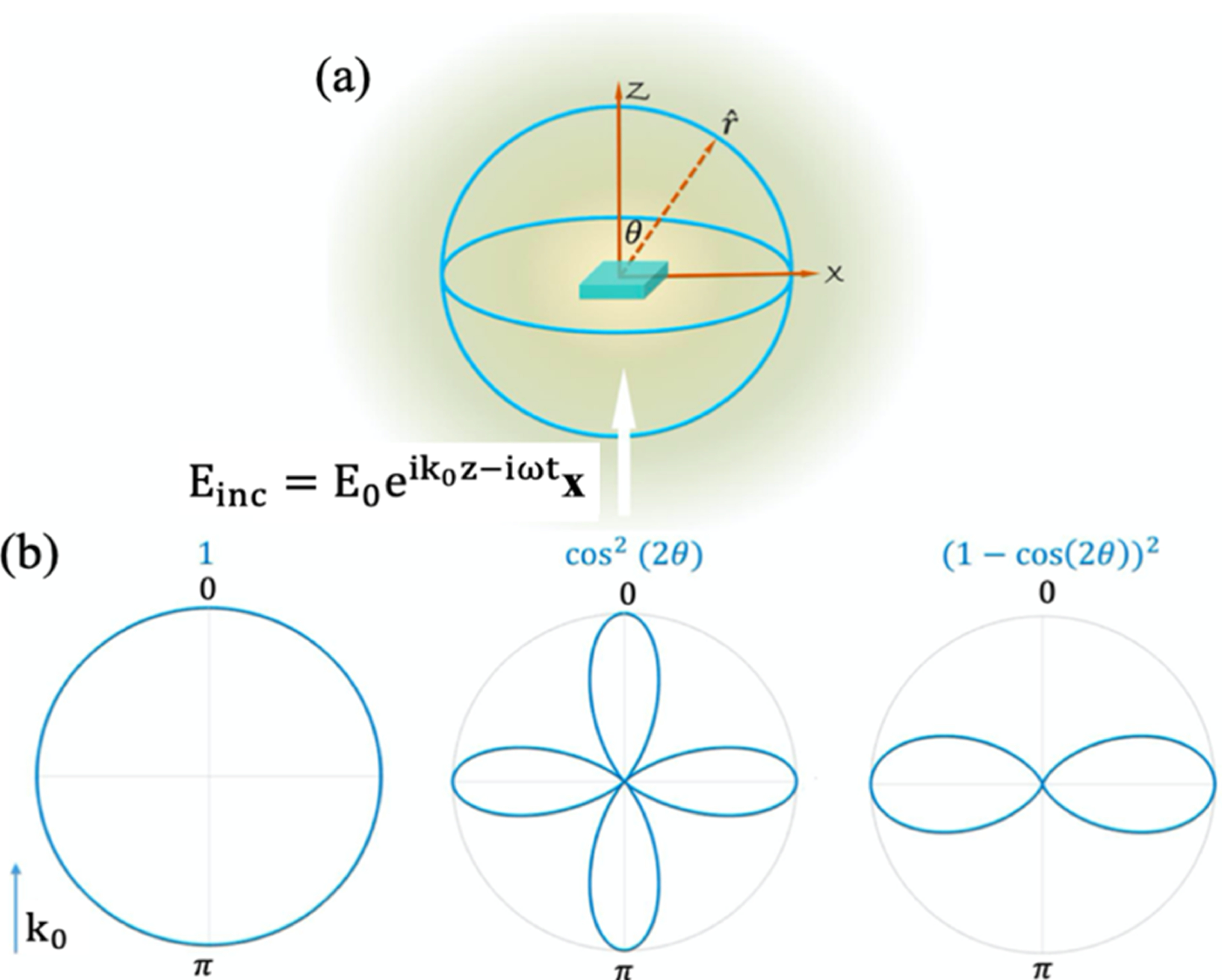}
\caption{\label{fig:Sch}(a) (a) Schematic graph of a single standalone square nanoplate embedded in air. $k_0$ is the wavevector in air. The color background is only for better visibility. (b) The polar angular weighting factor (Eq.2) of the ED, MQ and coherent sum of both when the phase differs by $\pi$ in $x-z$ plane. A normally incident plane wave illuminates the sample from below. The incident wave is polarized along $x$ direction and propagates along z direction. }
\end{figure}

We assume a standalone arbitrary subwavelength particle in air, the schematic can be seen in Fig.~\ref{fig:Sch}(a). The incident light travels along the $z$ axis and is polarized along the $x$ axis,  expressed as ${\textbf{E}_{inc}=E_0e^{ik_0z-i\omega t}\hat{\textbf{x}}}$, where $k_0$ is the free-space wave number and $\hat{\textbf{x}}$ is the unit vector along $x$ axis. The total scattered electric field along all directions considering up to quadrupole modes is defined as \cite{jackson1999classical,shamkhi2019transverse}
\begin{equation}
\begin{split}
\textbf{E}_{sca}=&\frac{k_0^2e^{i\textbf{k}\cdot\textbf{r}-i\omega t}}{4\pi r \epsilon_0}([\textbf{n}\times[\textbf{p}\times\textbf{n}]]
+\frac{1}{c}[\textbf{m}\times\textbf{n}]
\\&+\frac{ik_0}{6}[\textbf{n}\times[\textbf{n}\times(\textbf{Q}\cdot\textbf{n})]]+\frac{ik_0}{2c}[\textbf{n}\times(\textbf{M}\cdot\textbf{n})])
\end{split}
\end{equation}
where $\textbf{r}$ is the coordinate vector and its origin is placed at the center of nanoparticle. $\textbf{n}$ is the unit vector pointing from the origin to any coordinate position. $\epsilon_0$ is the vacuum permittivity. $\textbf{p}$, $\textbf{m}$, $\textbf{Q}$ and $\textbf{M}$ are ED moment, MD moment, EQ moment and MQ moment respectively. The expression of multipole modes is taken from Ref. \cite{terekhov2017multipolar} and the details are shown in Appendix A. In spherical coordinate, the differential scattering cross section as a function of the polarizabilities is expressed as

\begin{equation}
\frac{d\sigma}{d\Omega}(\theta)=[\frac{\pmb{\alpha_p}}{\epsilon_0}+\pmb{\alpha_m}cos\theta+\frac{k_0^2\pmb{\alpha_Q}}{12\epsilon_0}cos\theta+\frac{k_0^2}{4}\pmb{\alpha_M}cos(2\theta)]^2
\end{equation}
where $\pmb{\alpha_p}$, $\pmb{\alpha_m}$, $\pmb{\alpha_Q}$ and $\pmb{\alpha_M}$ are the polarizability of ED, MD, EQ and MQ respectively. $\theta$ is the polar angle in spherical coordinate as seen in Fig.~\ref{fig:Sch}. $\frac{d\sigma}{d\Omega}(\theta=0)$ corresponds to the forward scattering along $z$ direction and $\frac{d\sigma}{d\Omega}(\theta=\pi)$ corresponds to the backward scattering along $-z$ direction. 

As mentioned earlier, in the plane of incidence, the scattered electric field of the ED or MQ displays an even parity while the scattered field of the MD or EQ displays an odd parity \cite{liu2018generalized}. Transverse scattering, which is simultaneously zero forward and zero backward scattering, $\frac{d\sigma}{d\Omega}(\theta=0)=\frac{d\sigma}{d\Omega}(\theta=0)=0$, indicates a fundamental even parity. According to Eq.2, the scattering pattern manifests an even parity only $\pmb{\alpha_m}=0$ and $\pmb{\alpha_Q}=0$. Under this circumstance, transverse scattering can be achieved while satisfying two prerequisites. The first prerequisite is $\pmb{\alpha_p}=-\epsilon_0k_0^2\pmb{\alpha_M}/4$, under which condition the ED scattering destructively interferes with MQ scattering. The other prerequisite is that the polar angle weighting factor of polarizability also displays a transverse angular pattern in the plane of incidence. The polar angle weighting factors for the ED, MQ and the destructive interference, $(1-cos(2\theta))^2$, are shown in the bottom panel of Fig.~\ref{fig:Sch}, demonstrating the fulfilment of the second prerequisite.

Next we consider the more practical scenario of a metasurface in air. The backward scattering is interpreted as the reflection coefficient, $r$. The transmission coefficient, $t$, contains the forward scattering and the incident wave contribution. Under the illumination with a plane wave $\textbf{E}_{inc}=E_0e^{ik_0z-i\omega t}\textbf{x}$, ${r}$ and ${t}$ in Cartesian coordinates could be expressed as \cite{terekhov2019multipole}
\begin{equation}\label{eq:RTfull}
\begin{split}
&{r}=\frac{ik_0}{2E_0A\epsilon_0}(\textbf{p}_x-\frac{1}{c}\textbf{m}_y+\frac{ik_0}{6}\textbf{Q}_{xz}-\frac{ik_0}{2c}\textbf{M}_{yz})
\\&{t}=1+\frac{ik_0}{2E_0A\epsilon_0}(\textbf{p}_x+\frac{1}{c}\textbf{m}_y-\frac{ik_0}{6}\textbf{Q}_{xz}-\frac{ik_0}{2c}\textbf{M}_{yz})
\end{split}
\end{equation}
where $A$ is the area of a unit cell of the metasurface.  Multipolar modes are multipolar EM waves in origin with their inherent amplitude and phase along forward and backward directions. $\mathbf{p}_x$ refers to the total electric dipole, which includes the toroidal dipole contribution.

Similar to the single nanoparticle case, the multipolar EM wave displays a strictly even parity only if $|i\textbf{m}_y/c|$ and $|k_0\textbf{Q}_{xz}/6|$ are negligible compared with ED and MQ contributions. Then Eq.3 can be rewritten as ${r}=\frac{ik_0}{2E_0A\epsilon_0}(\textbf{p}_x-\frac{ik_0}{2c}\textbf{M}_{yz})$ and ${t}=1+\frac{ik_0}{2E_0A\epsilon_0}(\textbf{p}_x-\frac{ik_0}{2c}\textbf{M}_{yz})$. When the condition $r = 0$ is fulfilled, then $t = 1$, both forward and backward scattering from the metasurface disappear and transverse Kerker scattering is achieved. For the lossless case, the amplitude of incident light wave remains unperturbed. The metasurface behaves as non-existent or transparent, referred to as lattice invisibility effect or extraordinary transmission elsewhere \cite{terekhov2019multipole}. According to Eq.3, with negligible MD and EQ contribution, the generalized amplitude and phase conditions for transverse scattering are

\begin{equation}
\begin{split}
&  |i\textbf{m}_y/c|\approx0, |k_0\textbf{Q}_{xz}/6|\approx0
\\&|i\textbf{p}_{x}|=|\frac{k_0}{2c}\textbf{M}_{yz}|
\\&\varphi(i\textbf{p}_x)=\varphi(\frac{k_0}{2c}\textbf{M}_{yz})\pm(2n+1)\pi
\end{split}
\end{equation}
Additionally, perfect reflection can be achieved if $t = 0$, where the scattered EM waves destructively interfere with the transmitted incident electric field, in which case
\begin{equation}
\begin{split}
{r}=\frac{ik_0}{2E_0A\epsilon_0}(\textbf{p}_x-\frac{ik_0}{2c}\textbf{M}_{yz})=-1
\end{split}
\end{equation}
where for the lossless case, the incident wave is fully reflected and metasurface behaves as a perfect mirror and is referred to as a Huygens reflector.

\section{Metasurface invisibility for light at normal incidence}

\begin{figure}
\includegraphics[width=1.0\linewidth]{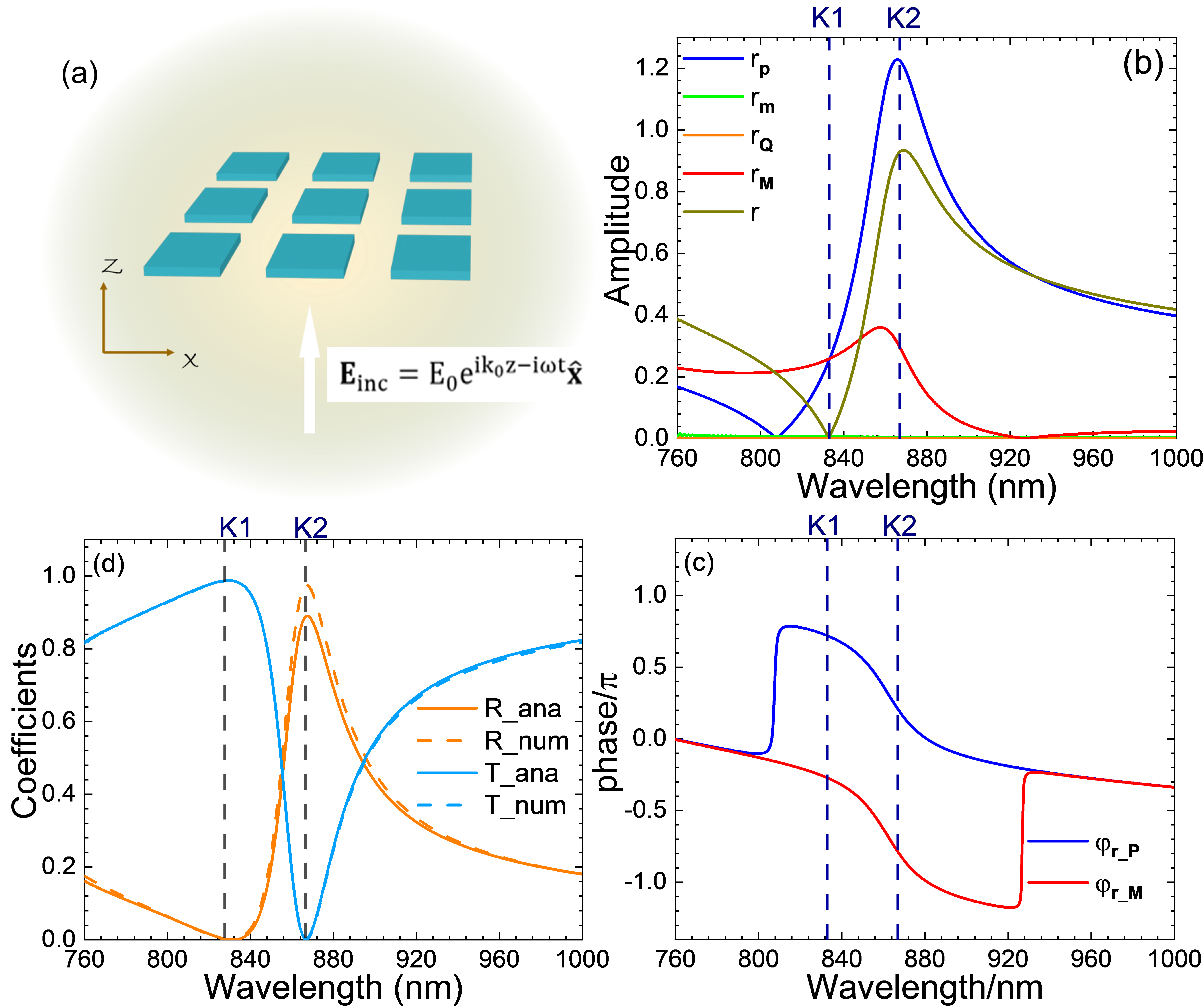}
\caption{\label{fig:APRT}(a) Schematic of the periodic Si square nanoplate metasurface embedded in air. The colour background is shown only  for better visibility of the nanostructure. The period is $D_x=D_y$= 750 nm, with L = 460 nm and H = 40 nm. A normally incident plane wave interacts with the metasurface. The incident wave, $\textbf{E}_{inc}=E_0e^{i\left(k_0z-\omega t\right)}\hat{\textbf{x}}$, is polarized along $x$ direction and propagates along $z$ direction. (b) Amplitude of the multipole moments from Eq.2, including $r_\textbf{p}$, $r_\textbf{m}$, $r_\textbf{Q}$, $r_\textbf{M}$ and $r$. (c) Phase profile of  $r_\mathbf{p}$ and $r_\mathbf{M}$. (d) Numerically simulated ($\_$num) and semi-analytically calculated ($\_$ana) reflection (R) and transmission (T) coefficients. Two wavelengths selected for detailed analysis are indicated by the vertical dashed lines, at K1 833 nm, K2 867 nm respectively.}
\end{figure}

Next we test the feasibility of the concept in a single nanoparticle or a metasurface with negligible MD and EQ contributions. An extremely-thin high refractive index metasurface, such as a Si square nanoplate metasurface, meets the prerequisites of negligible MD and EQ contributions. The much lower aspect ratio is crucial to suppress the MD and EQ contributions and is demonstrated by multipole decomposition following Ref. \cite{terekhov2017multipolar}. The conditions for negligible MD and EQ are determined by sweeping the geometrical aspect ratio, which is the ratio of the height, $H$, to the edge length of the square nanoplate, $L$. The aspect ratio is varied from 0.04 to 0.4, where the Mie-like mode is the scope of this study rather than  Fabry–P\'{e}rot-like modes within the metasurface \cite{rybin2017high,rybin2017switchable,bogdanov2019bound}. The dependence of amplitudes of multipolar modes on the height of the nanoplate or aspect ratio can be found in Supplemental Material Fig. S1. Additionally, the period along $x$, $D_x$, and $y$ directions, $D_y$, are also swept separately in order to explore the evolution of the ED and MQ modes, with details shown in Supplemental Material Fig. S2. The wavelength-dependent real and imaginary parts of refractive index and permittivity are obtained from fitting the experimental data in Ref.  \cite{aspnes1983dielectric}. All numerical simulations are performed using the Lumerical finite-difference time domain (FDTD) tool. Here we present, as an example. the results for the Si square nanoplate metasurface in air which has the edge length, $L$ = 460 nm, height $H$ = 40 nm, and period $D_x=D_y$ = 750 nm. The schematic representation can be seen in Fig.~\ref{fig:APRT}(a). An infinite periodic array of Si square nanoplates is illuminated by a normally-incident plane wave, $\textbf{E}_{inc}=E_0e^{i\left(k_0z-\omega t\right)}\hat{\textbf{x}}$, which propagates along $z$ direction and is polarized along $x$ direction. 

According to the standard expansion method in Ref. \cite{terekhov2017multipolar}, we consider the decomposition of the multipolar modes up to quadrupole modes by integrating the electric field over the square nanoplate volume. The decomposed multipole moments contributing to the reflection in Eq.3 are denoted as $r_{\textbf{p}}=ik_0\textbf{p}_x/(2E_0A\epsilon_0)$, $r_{\textbf{m}}=-ik_0\textbf{m}_y/(2cE_0A\epsilon_0)$,
$r_{\textbf{Q}}=-k_0^2\textbf{Q}_{xz}/(12E_0A\epsilon_0)$ and 
$r_{\textbf{M}}=k_0^2\textbf{M}_{yz}/(4cE_0A\epsilon_0)$. The corresponding amplitude and phase are presented in Fig.~\ref{fig:APRT}(b) and (c) respectively. It is clear in Fig.~\ref{fig:APRT}(b) that the amplitude of the MD and EQ is negligible. The dominant multipole moments are the ED and MQ.  The semi-analytically calculated and the directly numerically calculated reflection coefficient, $R=|r|^2$, and transmission coefficient, $T=|t|^2$, following Eq.3, are presented in Fig.~\ref{fig:APRT}(d), and are found to be in excellent agreement. This demonstrates that the multipole decomposition up to quadrupole modes is sufficient and consideration of higher order modes is not required. 

Special attention is paid to the wavelengths at 833 nm and 867 nm, denoted as K1 and K2, respectively, and shown by  as dash lines in Figs.~\ref{fig:APRT}(b), (c) and (d). At K1, ED and MQ share the same amplitude and differ only with a phase of $\approx\mathbf{\pi}$. According to Eq.3, $r = r_\mathbf{p}+r_\mathbf{M}$, the reflection coefficient  ${R=\left|r\right|^2=\left|r_\mathbf{p}\right|^2+\left|r_\mathbf{M}\right|^2+2{|r}_\mathbf{p}|{|r}_\mathbf{M}|cos(\varphi_{r_\mathbf{p}}-\varphi_{r_\mathbf{M}})\approx0}$. This explains the zero reflection coefficient at 833 nm. ED component, $r_\textbf{p}$ destructively interferes with MQ component $r_\textbf{M}$. As mentioned earlier, in this situation, $t=1$. The transmitted light has only the contribution from the incident wave without the scattering signal from the resonators in metasurface. This corresponds well with the near unity transmission at K1 as shown in  Fig.~\ref{fig:APRT}(d). Transverse Kerker scattering is expected at K1 according to generalized conditions as Eq.4.  Upon a closer look at K2 867 nm, ED and MQ do not have equal amplitude but also differ in phase by $\pi$. There is still destructive interference between ED and MQ according to the amplitude of $r_\textbf{p}$, $r_\textbf{M}$ and $r$ shown in Fig.~\ref{fig:APRT}(b) as $r < {r}_{p}$. However, ED and MQ cannot completely cancel each other at this wavelength due to unequal amplitudes. It is expected that the scattering pattern at K2 is symmetric in the forward and backward directions, displaying an even parity due to inherent even parity of ED and MQ modes.

\begin{figure}
\includegraphics[width=1.0\linewidth]{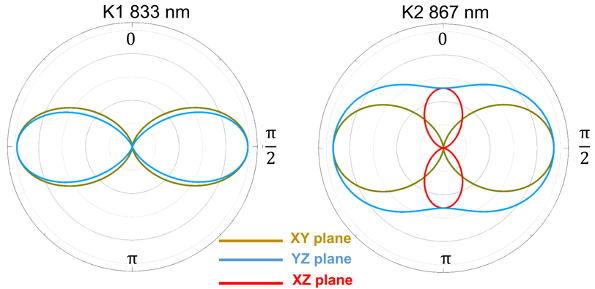}
\caption{\label{fig:pattern}The simulated far-field scattering power profile for one single resonator in the array, $|E|^2$ at K1 833 nm and K2 867 nm respectively. The angle 0 indicates the incident light propagation direction or the forward direction while $\pi$ indicates the backward direction for $x-z$ and $y-z$ plane. In $x-y$ plane, x direction points toward $\pi/2$.}
\end{figure}

\begin{figure*}
\includegraphics[width=0.8\linewidth]{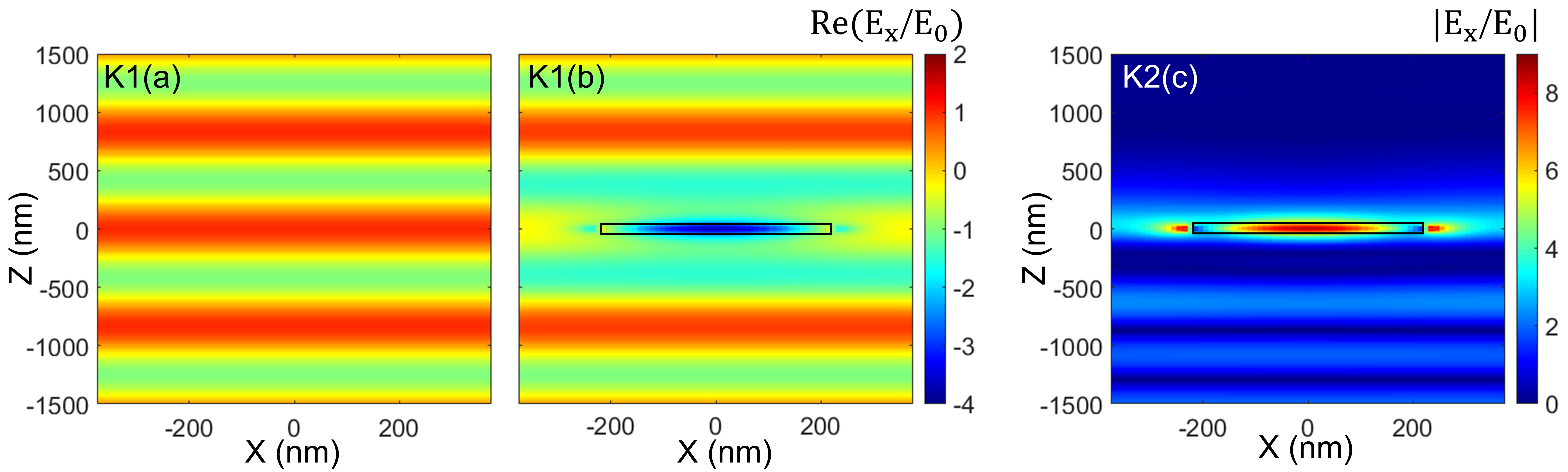}
\caption{\label{fig:map}Real part of the normalized scattered electric field, $\mathrm{Re(E_x/E_0)}$ distribution in in $x-z$ plane without (a) and with the square plate (b) at K1 = 833 nm. The square nanoplate resonator in one metasurface unit cell is shown by the black box. (c) The absolute value of the normalized scattered electric field, $\mathrm{|E_x/E_0|}$ distribution in the $x-z$ plane at K2 = 867 nm.}
\end{figure*}

To confirm further our expectations and also to provide an intuitive picture, the far field scattering intensity profile of one metasurface unit cell is calculated and presented in Fig.~\ref{fig:pattern} for both K1 833 nm and K2 867 nm. It is clear that the far-field scattering intensity pattern is symmetric in $x-y$, $y-z$ and $x-z$ planes, which is determined by the even parity of the ED and MQ modes. Notably, it is zero over the $x-z$ plane at K1 = 833 nm, including both the forward and backward directions, which manifests transverse scattering pattern. At K2 867 nm, it is clear that the far-field scattering intensity pattern is symmetric but forward and backward scattering still occur. The ED destructively interferes with MQ but full destructive interference is not achieved due to the unequal amplitudes. The real part of electric field at K1 and K2 wavelengths are presented in Fig.~\ref{fig:map} in the $x-z$ plane. As expected at K1 833 nm, the incident wave propagates along the $z$ direction and transmits through the metasurface almost without perturbation. The metasurface is rendered invisible, so-called lattice invisibility \cite{liu2019lattice}. At K2 867 nm, as expected, the incident wave, expressed as the absolute value of the amplitude ratio, $|E_x/E_0|$, is reflected back and the metasurface behaves as a perfect mirror. Note that as stated earlier, both forward and backward scattering remain at K2 867 nm, the coherent sum of ED and MQ destructively interferes with the transmitted incident EM wave, which results in near-zero transmission and near-unity reflection. This mechanism underpins the principle of the Si metasurface based perfect mirror.

\begin{figure*}
 \includegraphics[width=0.8\linewidth]{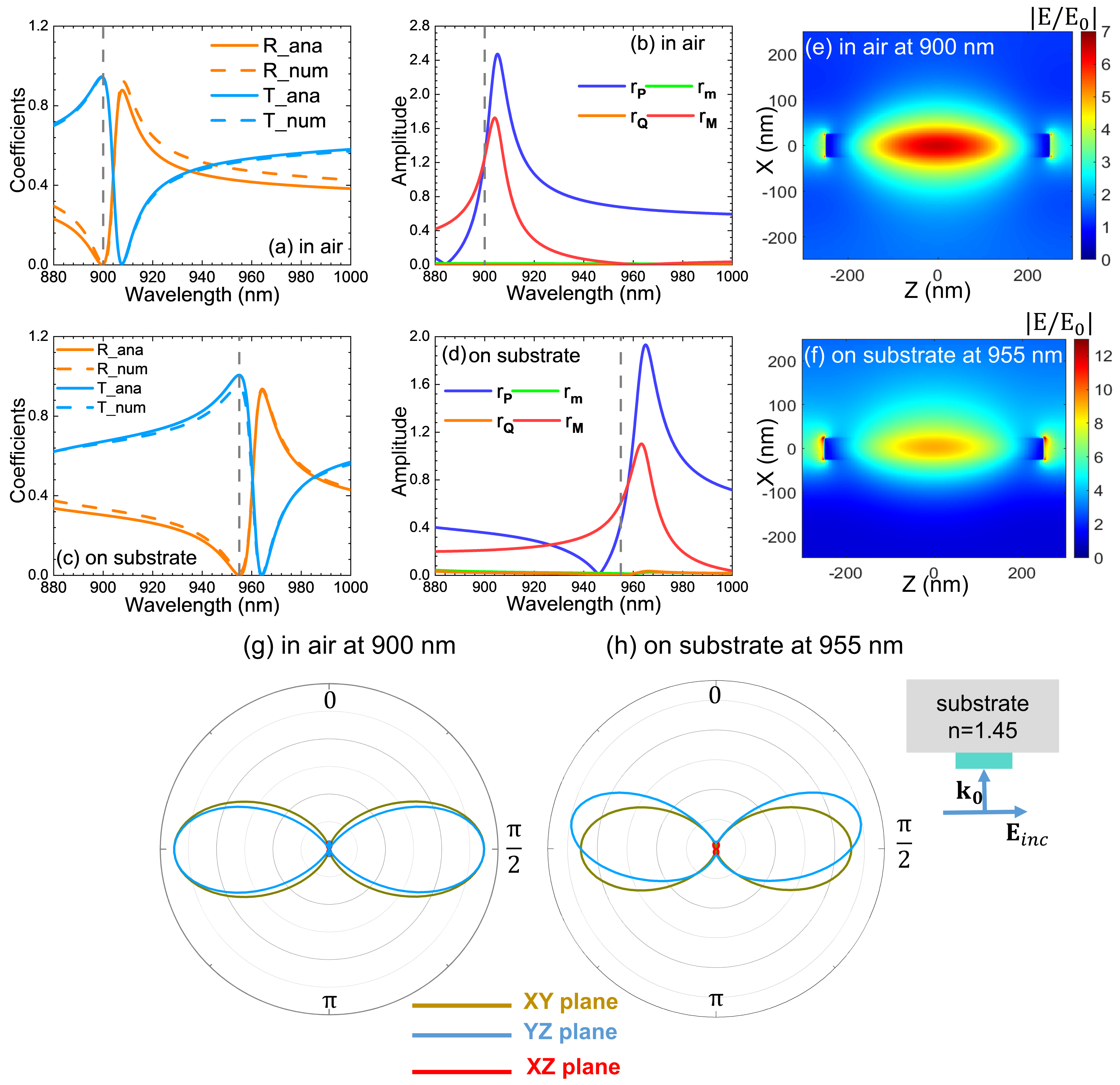}
 \caption{\label{fig:sub} (a) and (c) Numerically simulated ($\_$num) and semi-analytically ($\_$ana) calculated reflection (R) and transmission (T) coefficients of the Si square nanoplates ($L$ = 500 nm, $H$ = 50 nm, $D_x = D_y$ = 600 nm) in air and on a substrate, where the substrate has the refractive index 1.45. (b) and (d) The calculated amplitude of multipolar modes of the metasurface in air and on the substrate. The inspected wavelengths are shown as gray dash lines. (e) and (f) The electric field distribution in the $x-z$ plane and (g) and (h) the simulated far-field scattered power profile for one single resonator in the array at 900 nm in air and at 955 nm on the substrate, respectively. The metasurface in air or on the substrate is illuminated by normally incident plane wave propagating upwards, as shown in the schematic in (h). }
\end{figure*}

\section{Metasurface invisibility with TM polarizaed light at oblique incidence}
\begin{figure*}
\includegraphics[width=0.8\linewidth]{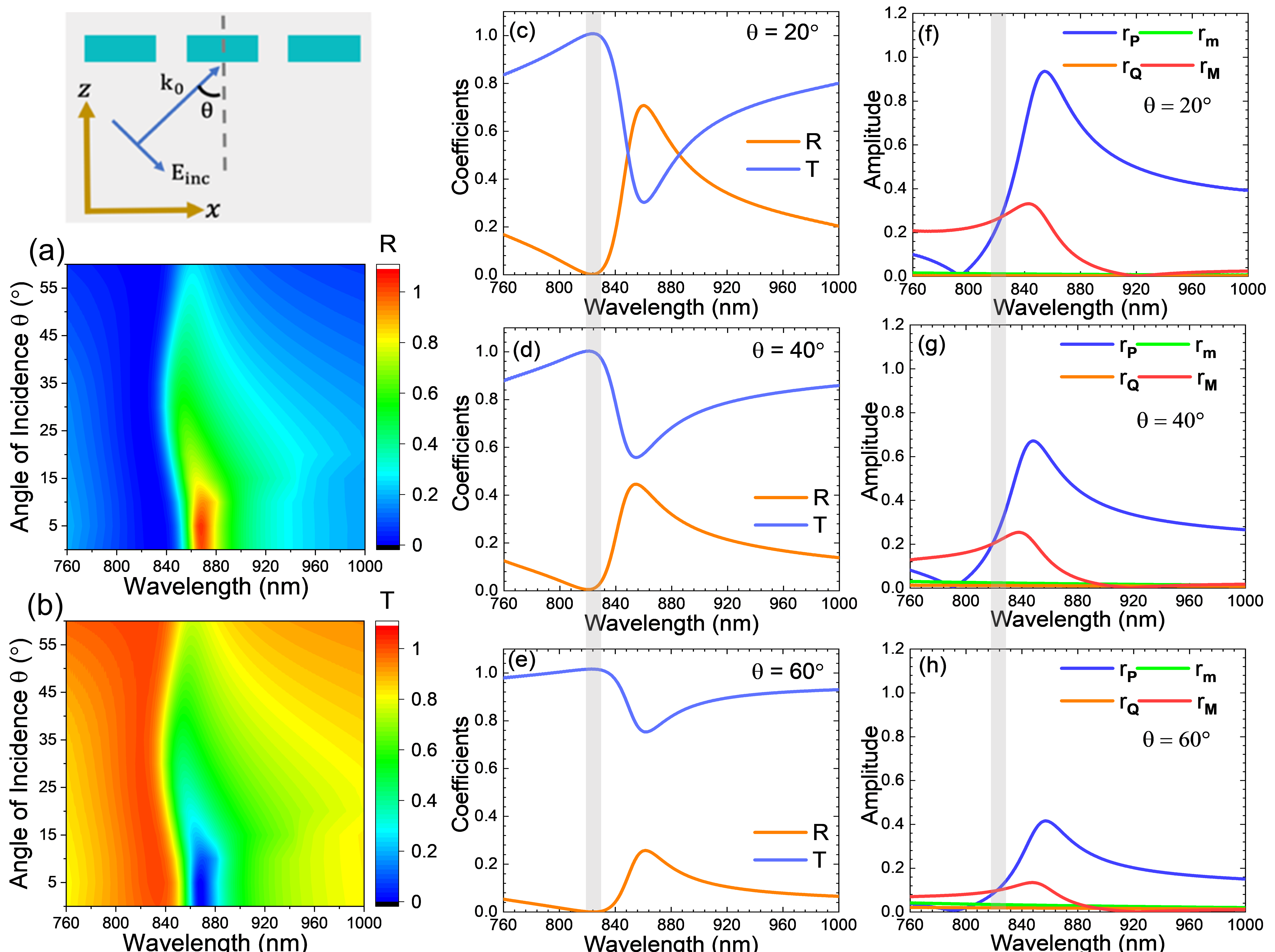}
 \caption{\label{fig:angle} The schematic showing transverse magnetic (TM) polarized oblique incidence light wave. $\theta$ is the angle between the wavevector $k_0$ and the $z$-direction, and is defined as the incident angle. (a) and (b) The contour plots of the numerically simulated reflection and transmission spectra for a metasurface with a TM-polarized plane wave incident at an oblique angle $\theta$ up to 60$^{\circ}$. (c), (d) and (e) The numerically simulated reflection and transmission spectra for the metasurface under the illumination with a TM polarized plane wave with an incident angle of $\theta$ = 20$^{\circ}$, $\theta$ = 40$^{\circ}$ and $\theta$ = 60$^{\circ}$, respectively. The shaded area illustrates the wavelength range where the metasurface is rendered nearly invisible or transparent.  (f), (g) and (h) The corresponding multipolar contributions generated by the TM plane wave incident at angles of $\theta$ = 20$^{\circ}$, $\theta$ = 40$^{\circ}$ and $\theta$ = 60$^{\circ}$, respectively.}
 \end{figure*}

With a view to practical applications, a metasurface with dominant ED and MQ on a substrate with low refractive index, such as refractive index 1.45, which matches the refractive index of glass, quartz and polymers, is now considered. Introducing the substrate will shift the spectral positions of resonant modes compared with that for a metasurface embedded in air medium, as can be seen Figs.~\ref{fig:sub}(b) and (d). Correspondingly the wavelengths at which invisible or near-perfect transparency occurs also shift, which is clearly seen in reflection and transmission spectra shown in Figs.~\ref{fig:sub}(a) and (c). The invisibility wavelength for the metasurface ($L$ = 500 nm, $H$ = 50 nm, $D_x$ = $D_y$ = 600 nm) shifts from 900 nm in air to 955 nm for the metasurface sitting on the substrate. Additionally, the substrate modifies the electric field distribution within the metasurface at the invisible wavelength, as can be seen from Figs.~\ref{fig:sub}(e) and (f). A larger electric field amplitude is obtained due to the presence of substrate. For the metasurface sitting on the substrate, the amplitudes of resonant modes are calculated from the electric field distribution within the metasurface sitting on the substrate rather than that in air. The reflection and transmission spectra obtained using the semi-analytical approach, according to Eq.3, agree well with the spectra calculated directly from the numerical simulation, as can be seen clearly from Fig.~\ref{fig:sub}(c). This demonstrates that the light reflected from substrate and Fabry-P\'erot multiple reflections within the ultra-thin metasurface can be considered as negligible \cite{babicheva2017reflection}. Additionally, as seen in Fig.~\ref{fig:sub}(g), the scattering profile at 900 nm, for the metasurface in air, shows a perfect transverse scattering pattern, where the scattering lobe is symmetric and mainly in transverse directions. However,  the scattering pattern is modified due to the existence of substrate as seen in Fig.~\ref{fig:sub}(h), where an asymmetric pattern in $x-z$ and $y-z$ planes is observed. There is slight tilt of the side lobes, though it is less than has been observed in the case of employing the coherent interplay of ED, MD, EQ and MQ to achieve transverse scattering, as reported for example in Ref. \cite{shamkhi2019transverse}. The approach of using only the coherent interplay of ED and MQ appears to be less affected by the substrate, with the scattering mainly confined to  the transverse plane.

In this section, we explore how the invisibility evolves under excitation with light at oblique angles of incidence for the metasurface in air. Oblique plane wave incidence changes the spectral positions of the resonant multipoles \cite{bogdanov2019bound}.  Transverse magnetic (TM) oblique wave incidence corresponds to a change in $D_x$ while transverse electric oblique wave incidence corresponds to a change in $D_y$.  As mentioned earlier, the effect of lattice constant or period along $x$ and $y$ directions, respectively, can be found in Supplemental Material Fig.S2. For the metasurface illuminated by a normally-incident plane wave polarized along the $x$-axis, the spectral positions of resonant ED and MQ modes are not significantly affected by the $D_x$ but gradually red-shift with increasing $D_y$. TM oblique incidence corresponds to changes of $D_x$ \cite{babicheva2017reflection}, therefore, invisibility could be expected under oblique TM incidence. In order to further explore the possibility of wide-angle invisibility,  the Broadband Fixed Angle Source Technique (BFAST) is used for broadband angle-resolved simulations using Lumerical FDTD  \cite{liang2020bound,babicheva2018lattice,gao2019ultraviolet,liang2013wideband}.

The contour plots of the numerically simulated angle-resolved reflection and transmission spectra can be seen in Figs.~\ref{fig:angle}(a) and (b). It is clear that the invisibility, or transparency, still remains at oblique angle incidence although with a slight spectra shift. The wavelength range, or transparency wavelength window, is illustrated as the shaded area in Figs.~\ref{fig:angle}(c), (d) and (e), where near-perfect transparency exists at all angles. To probe more the underlying physics, multipolar decomposition analysis for various angles of incidence for a TM polarized plane wave, including 20$^{\circ}$, 40$^{\circ}$, 60$^{\circ}$, is undertaken and shown in Figs.~\ref{fig:angle}(f), (g) and (h). With increasing angle of incidence, $\theta$, where the incident light polarization can be decomposed into $z$ and $x$ polarized light waves, the amplitude of the $x$-polarized wave becomes effectively $|E_{inc}cos\theta|$, and therefore the amplitude of ED and MQ decrease, which is clearly seen in Figs.~\ref{fig:angle}(f),(g) and (h). However, overlapping resonant modes is still achieved and the spectral position of the overlap does not shift significantly, as can be seen by shaded area. The invisibility studied in this work is driven by transverse Kerker scattering, or destructive interference of ED and MQ, therefore the equal amplitude, as seen in Figs.~\ref{fig:angle}(f), (g) and (h), and coherence properties of ED and MQ guarantees the wide-angle invisibility or transparency.

\section{Conclusion}

To conclude, starting from a general description the resonant multipolar modes for a single arbitrary subwavelength nanoparticle and the properties of transverse Kerker scattering are discussed. The transverse Kerker scattering field and intensity exhibit an even parity in the forward and backward directions. The description was then extended to consider  metasurface, in which rather than employing the coherent interplay of the ED, MD, EQ and MQ to achieve transverse scattering, only the ED and MQ multipoles are employed. The proposed route is numerically realized in an ultra-thin Si square nanoplate metasurface in air and could be extend to any extremely-thin arbitrary shape nanoparticle, where  a low aspect ratio guarantees negligible MD and EQ contributions. It is shown that the proposed configuration is valid for a metasurface embedded in a homogeneous medium or on a substrate with low refractive index, such as polymer, glass or quartz. Additionally, it is demonstrated that due to the extremely-thin layer thickness, our proposed metasurface shows robust TM incident angle independence where the invisibility or near-perfect transparency can be achieved, although there is slight spectra position shifts. Our study can inspire new Huygens metasurface design with ultra-thin nanostructures and spur further experimental investigations for perfect mirrors, perfect transmission and other invisibility applications, as well as exploitation of the transverse scattered fields.

\begin{acknowledgments}
We wish to acknowledge the support of Science Foundation Ireland (SFI) under Grant Number 16/IA/4550.
\end{acknowledgments}

\appendix

\section{Appendix}
The expression of multipolar modes in Cartesian Coordinate is:
\begin{equation}
\begin{split}
&\textbf{p}=\int \epsilon_0(\epsilon_{Si}-1)\textbf{E}(\textbf{r})d\textbf{r}
\\&\textbf{T}=\frac{-i\omega}{10c}\int\epsilon_0(\epsilon_{Si}-1)[(\textbf{r}\cdot\textbf{E}(\textbf{r}))\textbf{r}-2\textbf{r}^2\textbf{E}(\textbf{r})]d\textbf{r}
\\&\textbf{m}=-\frac{i\omega}{2}\int\epsilon_0(\epsilon_{Si}-1)[\textbf{r}\times\textbf{E}(\textbf{r})]d\textbf{r}
\\&\textbf{Q}=3\int\epsilon_0(\epsilon_{Si}-1) [\textbf{r}\textbf{E}(\textbf{r})+\textbf{E}(\textbf{r})\textbf{r}-\frac{2}{3}(\textbf{r}\cdot\textbf{E}(\textbf{r}))\hat{U}]d\textbf{r}
\\&\textbf{M}=\frac{\omega}{3i}\int\epsilon_0(\epsilon_{Si}-1)[(\textbf{r}\times\textbf{E}(\textbf{r}))\textbf{r}+\textbf{r}(\textbf{r}\times\textbf{E}(\textbf{r}))]d\textbf{r}
\end{split}
\end{equation}
where $\textbf{r}$ is the coordinate vector with its origin placed at the center of square nanoplate. $\textbf{E}(\textbf{r})$ is the total electric field inside the nanoplate at different position. $\epsilon_0$ is the vacuum permittivity; $\epsilon_{Si}$ is the relative dielectric permittivity of the Si particle. $c$ is the light speed in vacuum; $\hat{U}$ is the 3$\times$3 unity tensor; \textbf{p}, \textbf{T}, \textbf{m}, \textbf{Q} and \textbf{M} are the moments of ED, TD, MD, electric quadrupole (EQ) and MQ respectively.

\nocite{*}

\bibliography{apssamp}
\textbf{Supplementary Material}

\subsection{Si square nanoplate height, H or aspect ratio effect }
\renewcommand\thefigure{S\arabic{figure}}
\begin{figure*}
\setcounter{figure}{0}
\includegraphics[width=0.8\linewidth]{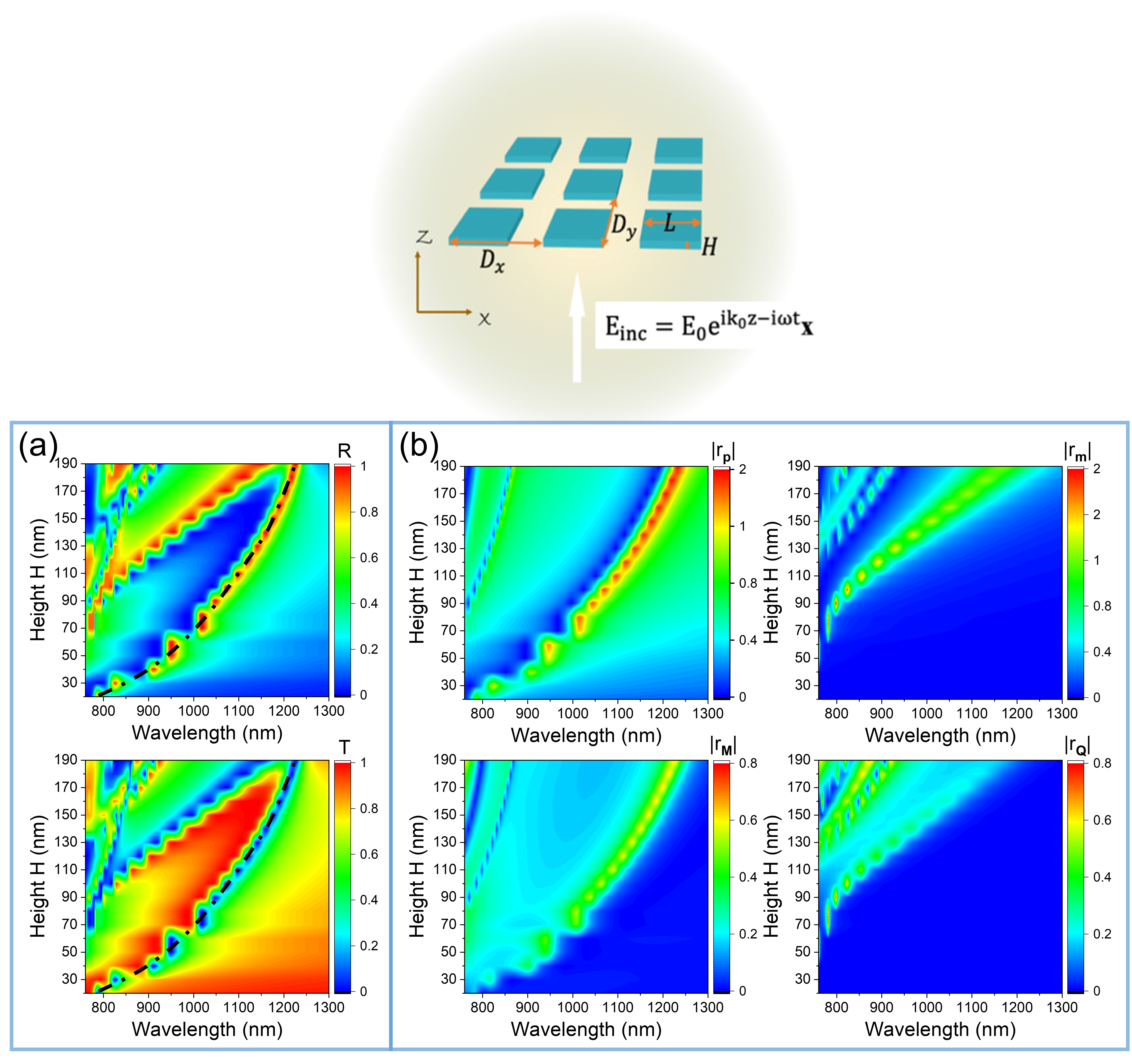}
\caption{\label{fig:height}Schematic graph of Si square nanoplate metasurface in air: $D_x = D_y$ = 750 nm, $L$ = 460 nm. (a) The simulated Reflection (R) and Transmission (T) spectra of the metasurface as a function of the height $H$, ranging from 20 nm to 190 nm. The black dash line is the guide to the eyes of the spectra changes as varying height $H$. (b) The calculated corresponding multipolar contributions to the reflection and transmission spectra. }
\end{figure*}

The reflection coefficient, $R$ and transmission coefficient, $T$ are shown in Fig. S1(a). The  calculated amplitude of corresponding multipolar modes are shown in Fig. S1 (b) according to the standard expansion method \cite{terekhov2017multipolar}. The available resonant multipolar modes, including the electric dipole (ED), $\textbf{p}$, magnetic dipole (MD), $\textbf{m}$, electric quadrupole (EQ), $\textbf{Q}$ as well as magnetic quadrupole (MQ), $\textbf{M}$, gradually shift to the longer wavelength as the height or the aspect ratio increases. It is clear that EQ and MD are negligible when the height is below 60 nm, or the aspect ratio is lower than 0.08. The ED and MQ dominate and determine the reflection and transmission spectra, and the transverse Kerker scattering mediates the lattice invisibility or transparency.

\subsection{Lattice effect}

Fig. S2 shows the contour plot of the simulated reflection and transmission spectra of metasurface with varying $D_x$ in (a) and $D_y$ in (b), respectively. When the metasurface is illuminated by a normally-incident light wave which polarizes along $x$ axis, the reflection/transmission spectra varies significantly and depends on the period along $y$ direction $D_y$, while it is relatively independent of the period in the $x$ direction, $D_x$.  The available ED and MQ modes do not shift with $D_x$ period but red-shift gradually with increasing $D_y$ or  inter-particle distance along the $y$ direction. More details could be seen in Fig. S2(c) as an example. These results agree with the results summarized in Ref. \cite{babicheva2018lattice}. 

\renewcommand\thefigure{S\arabic{figure}}
\begin{figure*}
\includegraphics[width=0.7\linewidth]{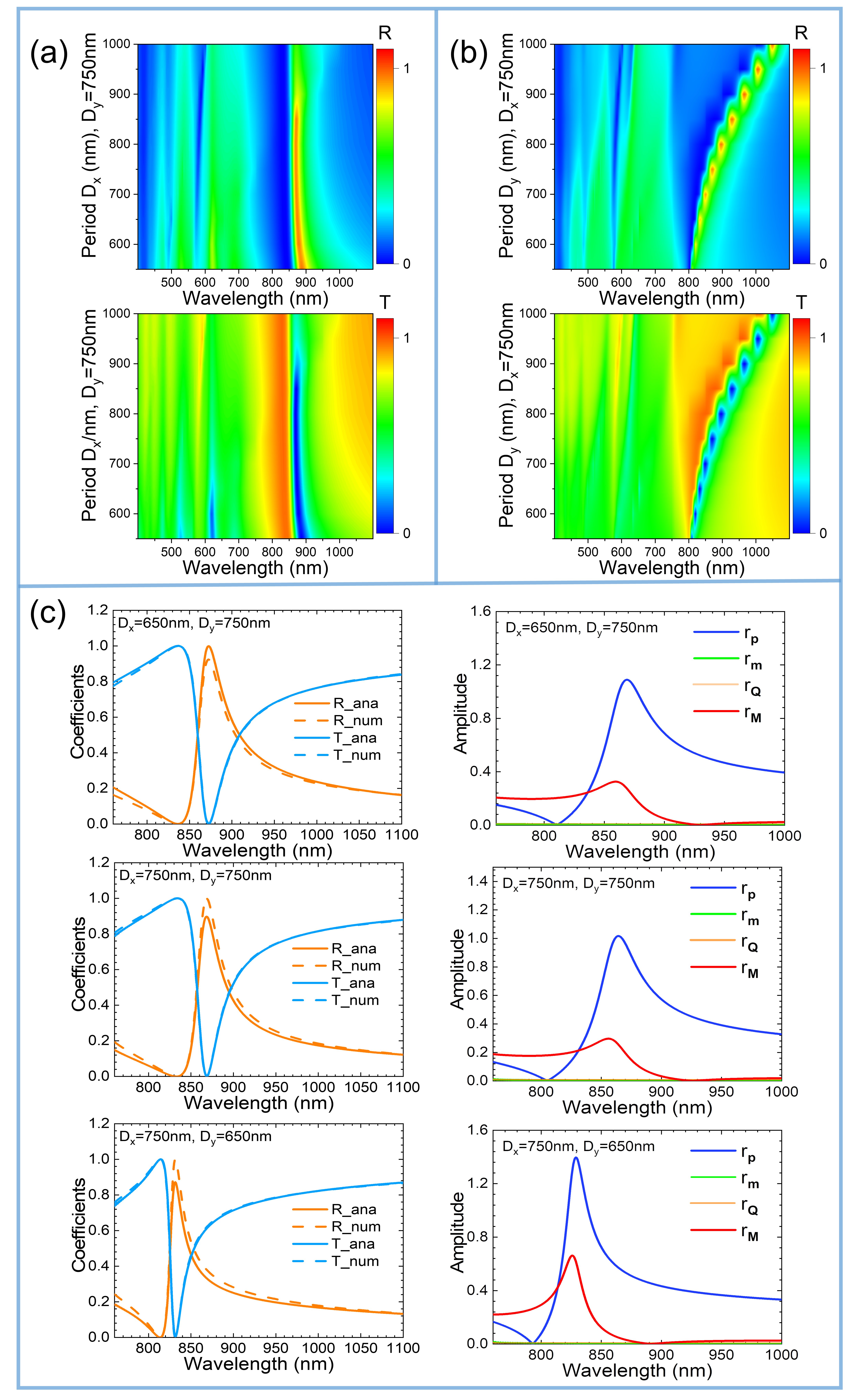}
\caption{\label{fig:period}The metasurface is illuminated by normally incident light which polarized along x axis. The numerically simulated reflection ($R$) and transmission ($T$) spectra of the metasurface for (a) $D_y$ = 750 nm with varying $D_x$ and (b) $D_x$ = 750 nm with varying $D_y$. (c) The numerically simulated and analytically calculated reflection and transmission coefficient and multipole decomposition for ($D_x$, $D_y$)=(650nm, 750nm), (750nm, 750nm) and (750nm, 650nm), respectively.}
\end{figure*}

\end{document}